\documentclass[aps,pre,twocolumn,superscriptaddress,showpacs]{revtex4-1}
\usepackage{graphicx}
\usepackage{amsmath}
\usepackage{natbib}
\usepackage[usenames]{color}
\usepackage{soul}

\usepackage[normalem]{ulem}

\usepackage{amsfonts}
\usepackage{dcolumn}        
\usepackage{amssymb}
\usepackage{bm,amssymb}

\usepackage{bm,fancybox}                	     

\begin{document}

\newcommand{\bea}{\begin{eqnarray}}
\newcommand{\eea}{  \end{eqnarray}}
\newcommand{\bit}{\begin{itemize}}
\newcommand{\eit}{  \end{itemize}}

\newcommand{\be}{\begin{equation}}
\newcommand{\ee}{\end{equation}}
\newcommand{\ra}{\rangle}
\newcommand{\la}{\langle}
\newcommand{\U}{\widetilde{U}}

\def\bra#1{{\langle#1|}}
\def\ket#1{{|#1\rangle}}
\def\bracket#1#2{{\langle#1|#2\rangle}}
\def\inner#1#2{{\langle#1|#2\rangle}}
\def\expect#1{{\langle#1\rangle}}
\def\e{{\rm e}}
\def\proj{{\hat{\cal P}}}
\def\tr{{\rm Tr}}
\def\H{{\hat H}}
\def\Hdag{{\hat H}^\dagger}
\def\Lop{{\cal L}}
\def\Ehat{{\hat E}}
\def\Edag{{\hat E}^\dagger}
\def\Shat{\hat{S}}
\def\Sdag{{\hat S}^\dagger}
\def\Ahat{{\hat A}}
\def\Adag{{\hat A}^\dagger}
\def\U{{\hat U}}
\def\Udag{{\hat U}^\dagger}
\def\Zhat{{\hat Z}}
\def\Phat{{\hat P}}
\def\Op{{\hat O}}
\def\id{{\hat I}}
\def\x{{\hat x}}
\def\P{{\hat P}}
\def\Px{\proj_x}
\def\Pr{\proj_{R}}
\def\Pl{\proj_{L}}


\title{Lagrangian descriptors for open maps}

\author{Gabriel G. Carlo}
\email[E--mail address: ]{carlo@tandar.cnea.gov.ar}
\affiliation{Comisi\'on Nacional de Energ\'{\i}a At\'omica, CONICET, 
Departamento de F\'{\i}sica, 
Av.~del Libertador 8250, 1429 Buenos Aires, Argentina}

\author{F. Borondo}
 \email[E--mail address: ]{f.borondo@uam.es}
\affiliation{Departamento de Qu\'{\i}mica, 
 Universidad Aut\'onoma de Madrid,
 Cantoblanco, 28049--Madrid, Spain}
\affiliation{Instituto de Ciencias Matem\'aticas (ICMAT),
 Cantoblanco, 28049--Madrid, Spain}
\date{\today}

\begin{abstract}

We adapt the concept of Lagrangian descriptors, which have been recently 
introduced as efficient indicators of phase space structures in chaotic systems, 
to unveil the key features of open maps. We apply them to the open tribaker map, 
a paradigmatic example not only in classical but also in quantum chaos. 
Our definition allows to identify in a very simple way the inner 
structure of the chaotic repeller, which is the fundamental invariant set 
that governs the dynamics of this system. The homoclinic tangles of periodic 
orbits (POs) that belong to this set are clearly found. 
This could also have important consequences for chaotic scattering 
and in the development of the semiclassical theory of short POs for open systems. 

\end{abstract}
\maketitle

\section{Introduction}
 \label{sec:intro}

Lagrangian descriptors (LDs) \cite{LD1} are a recently introduced classical measure 
which has proven to be very useful for the identification of the stable and unstable 
manifolds of chaotic systems \cite{LD2,LD3,LD4}.

They have also been applied to unveiling the chaotic structure in phase space 
of molecules, in particular the LiCN one \cite{LD5} which is described 
by a realistic potential in two and three dimensions. 
This study has demonstrated the ability of LDs to overcome the difficulty posed by 
higher dimensionality to other methods such as obtaining a Poincar\'e surface of section.
Also, LDs have been successfully implemented \cite{LD6} within the so-called 
geometric transition state theory to the identification of recrossing-free
dividing surfaces, thus helping in the computation
of chemical reaction rates, and the reactive islands that account for 
nonstatistical behavior in chemical reactions \cite{LD7}.
Moreover, the concept has been adapted to discrete dynamical systems like bidimensional 
area preserving maps \cite{LD8}, under the name of discrete LDs. 
In this work the singular sets of LDs have been associated to the invariant manifolds 
of some prototypical maps and a chaotic saddle has been identified. 
As an interesting example of this method, LDs have been successfully applied to the 
Arnold's cat map \cite{LD9} and its invariant manifolds have been easily described.  
 
In this paper, we introduce a measure that is closely related to 
the original LDs but modified in such a way that makes it specially suitable to 
uncover the structure embedded in the repellers that characterize open systems \cite{Clauss}. 
Instead of the chaotic saddle formed by the union of the stable and unstable manifold, 
the prevailing object in the phase space of scattering or projectively open systems 
which are not area preserving is the set of non escaping trajectories in the past and future. 
For that reason the LDs concept needs to be adapted to reveal the intersection of these 
manifolds. 
 
We focus in open maps on the torus and we take the open tribaker map as a 
paradigmatic benchmark example. The simplest way to make a map open 
consists in eliminating the trajectories going through an 'opening' in phase space.
For long times, this leaves just a repeller, which is a fractal invariant set. 
Nevertheless, reflection mechanisms at the boundaries are 
usually more complicated than a complete opening \cite{WiersigFresnel}, 
and they may exhibit many interesting mathematical consequences \cite{WiersigR}. 
This leads us to consider in this work a function depending on a reflectivity $R$, 
which rules the way in which the classical trajectories arriving at the opening are 
only partially reflected. 
We study two cases, namely a (discontinuous) constant 
reflectivity function, and another of the Fermi-Dirac type
that makes the boundaries of the opening smooth. 

We have found that our modified LDs are very good indicators for the homoclinic 
tangles associated to POs, which are not easy to describe in general. 
In particular, the short POs and their homoclinic associates are readily 
localized with this measure. 
From the classical point of view, this is potentially very useful 
in the theory of chaotic scattering \cite{ChaoticScattering,WigginsBook}. 
Moreover, it has applicability in the semiclassical theory of short 
POs \cite{art0,art1,art2,art3,art4}, and the study of the morphology of 
chaotic eigenstates \cite{art5}.  
 
This is how the paper is organized: 
In Sec.~\ref{sec:LDsforOM} we define the Lagrangian descriptors for open maps, 
and describe the open tribaker map together with its main properties. 
In Sec.~\ref{sec:results} we apply this definition to uncover the underlying 
structure of classical repellers, explaining our findings by using symbolic dynamics. 
Finally, our concluding remarks are presented in Sec.~\ref{sec:conclusions}.

\section{Lagrangian descriptors for open maps}
 \label{sec:LDsforOM}

Essential properties of generic dynamical systems are well described 
by maps \cite{Ozorio 1994,Hannay 1980,Espositi 2005}, 
and then they have been widely used as prototypical models of chaos. 
If we consider a bidimensional phase space (canonical variables $q$ and $p$) 
with an opening, i.e.~a region through which trajectories can escape, 
we have open maps. 
These kind of transformations of the 2-torus can be used to model chaotic 
scattering \cite{ChaoticScattering} and microlasers \cite{microlasers}, for example.  
The main invariant set that rules their properties is the so-called repeller, 
which has fractal dimension, and it is formed by the 
intersection of the forward and backwards trapped sets. 
These, in turn, are made of the trajectories that never escape 
either in the past or in the future, respectively. 
The repeller is usually characterized by means of 
a measure $\mu(X_i)$ at each phase space region $X_i$ determined by the average 
intensity $I_t$ when $t \rightarrow \infty$ of a number $N_{\rm ic}$ of initial 
conditions randomly chosen inside $X_i$. 
The initial intensity is $I_0=1$ for each trajectory, and it is decreased 
as $I_{t+1}=F_R(q,p) I_t$ each time it hits the opening \cite{Noeckel} 
($F_R(q,p)$ is the reflectivity function to be defined in what follows). 
A finite time approximation to the measure for $X_i$ can be defined by
$\mu_{t,i}^{b}=\langle I_{t,i} \rangle/\sum_i \langle I_{t,i} \rangle$
where the average is performed over the initial conditions in the given 
phase space region. 
This is the finite time backwards trapped set of open maps, and if evolved backwards 
$\mu_{t,i}^{f}$ is obtained, which is the forward trapped set. 
The intersection $\mu_{t,i}^{b} \cap \mu_{t,i}^{f}$ is the finite time repeller $\mu_{t,i}$.
 
However, this quantity does not give information regarding the inner structure 
of the invariant set. 
In order to throw light on this, we modify the original definition of discrete LDs 
for area preserving maps. 
If we consider a trajectory $\{q_t,p_t\}_{t=-T}^{t=T}$, where $t\in \mathbb{N}$, 
discrete LDs were defined in \cite{LD8} as 
\begin{equation}
 {\rm LD}_a=\sum_{t=-T}̣̣^{t=T-1} |q_{t+1}-q_t|^a+|p_{t+1}-p_t|^a, 
\end{equation}
with $a \leq 1$. 
As the repeller is the intersection of the backwards and forward trapped sets, 
we define now the LDs for open maps as 
\begin{equation}
  \begin{array}{lc}
 {\rm LDO}_a=\sum_{t=-T}̣̣^{t=-1} (|q_{t+1}-q_t|^a+|p_{t+1}-p_t|^a) I_t \times \\ 
 \sum_{t=0}̣̣^{t=T-1} (|q_{t+1}-q_t|^a+|p_{t+1}-p_t|^a) I_t,
   \end{array}
 \label{Eq:LDO}
\end{equation}
with $a=-0.3$ throughout this work (we drop the subscript $a$ in the following). 
Notice that we require $a<0$ in order to provide a direct comparison with the measure $\mu$. 
We take a $3^5 \times 3^5$ square grid on the torus with $N_{\rm ic}=10^3$ at each region 
$X_i$ so defined; also we normalize the LDOs to $1$.

The paradigmatic tribaker map is the model chosen for our studies since this is one 
of the simplest chaotic maps which can be described by a ternary Bernoulli shift. 
Moreover, the openings can be chosen to follow the stable and unstable manifolds. 
The open map is then simply the composition of the closed tribaker transformation 
(area-preserving, uniformly hyperbolic, piecewise-linear and invertible map with
Lyapunov exponent $\lambda=\ln{3}$)
\begin{equation}
\mathcal B(q,p)=\left\{
  \begin{array}{lc}
  (3q,p/3) & \mbox{if } 0\leq q<1/3 \\
  (3q-1,(p+1)/3) & \mbox{if } 1/3\leq q<2/3\\
  (3q-2,(p+2)/3) & \mbox{if } 2/3\leq q<1\\
  \end{array}\right\}
\end{equation}
with the selected opening. 
Notice that the opening mechanisms are not always as simple as a constant 
discontinuous function of $q$ and $p$ \cite{art4}. 
In fact, partially open maps are those in which the opening reflects some of 
the trajectories that arrive at it \cite{art5}. 
In this paper we study two different functions of the phase space $F_R(q,p)$, 
where $R$ is the parameter that determines a transition between a given minimum 
amount of reflection $R=0$ and the closed map ($R=1$). 
We define the opening region as the domain $1/3<q<2/3$ of the 
reflectivity function $F_R$. 
We take a constant function given by the value of $R$ in the opening 
and $1$ elsewhere. 
In this case, we obtain a complete opening for $R=0$; 
in all the other cases some amount of the incoming orbits is reflected. 
The other reflectivity that is considered in this work is given by
\begin{equation}
F_R(q,p)=\left\{
  \begin{array}{lc}
  (1-R)/(1+\exp(-A(q-B)))+R  \\ 
   \mbox{if } q>1/2 \\ 
  (1-R)/(1+\exp(-A((1-q)-B)))+R  \\ 
  \mbox{if } q<1/2,\\ 
 \end{array}\right.
 \label{reflectivity1}
  \end{equation}
which consists of a Fermi-Dirac type step function. 
We fix $A=120$ and $B=0.63$, that gives a value of approximately $1$ at $q=1/3$ and $q=2/3$, 
and the minimum value $R$ at the middle of the opening. 
This function represents a smoothing of the hard step considered in the previous case. 

The symbolic dynamics associated to the map action is very simple, 
being given by a Bernoulli shift in the ternary representation of 
$q=0.\epsilon_0 \epsilon_1 \epsilon_2 \ldots$ and 
$p=0.\epsilon_{-1} \epsilon_{-2} \epsilon_{-3} \ldots$ where $\epsilon_i=0,1,2$, as
\begin{equation}\label{eq.shift}
(p \vert q)=...\epsilon_{-2}\epsilon_{-1}.\epsilon_{0}\epsilon_{1}...
\xrightarrow{\mathcal{B}}
(p^\prime \vert q^\prime)=...\epsilon_{-2}\epsilon_{-1}\epsilon_{0}.\epsilon_{1}...
\end{equation}
Hence, applying the map simply implies to move the decimal point one position to the right. 
For the opening region that we have chosen and in the case of the discontinuous function,
it is clear that the equivalent in ternary notation can be obtained by using open symbols 
$\tilde{\epsilon}$ with forbidden value 1 ($\tilde{\epsilon}=0,2$). 
We consider only this possibility in order to study all reflectivity situations (and of course all  
symbols for the closed map case). 
One of the advantages of having such a simple symbolic dynamics 
(which unfortunately is not the usual situation) is that POs and their associated 
homoclinic tangles can be computed very easily. 
These sets are of the utmost relevance for chaotic scattering for example, but also 
for the semiclassical theory of short POs \cite{Vergini}. 
In this case, POs are simply given by an infinite repetition of a string of symbols 
$\nu=\epsilon_{0}...\epsilon_{L}$, where $L$ is the period. 
The homoclinic tangle of $\nu$, which is formed by the orbits that belong to both the stable and 
unstable manifolds of the PO, can be approximated at short times by strings 
of the form $\nu_{H}^{e}=\nu...\nu \epsilon^{e} \nu...\nu$,
where the orbit $\nu$ is repeated $e$ times at each end, 
and the homoclinic excursions strings $\epsilon^{e}$ go from length $1$ up to $e \in \mathbb{N}$.

%
\begin{figure}
\includegraphics[width=8.5cm]{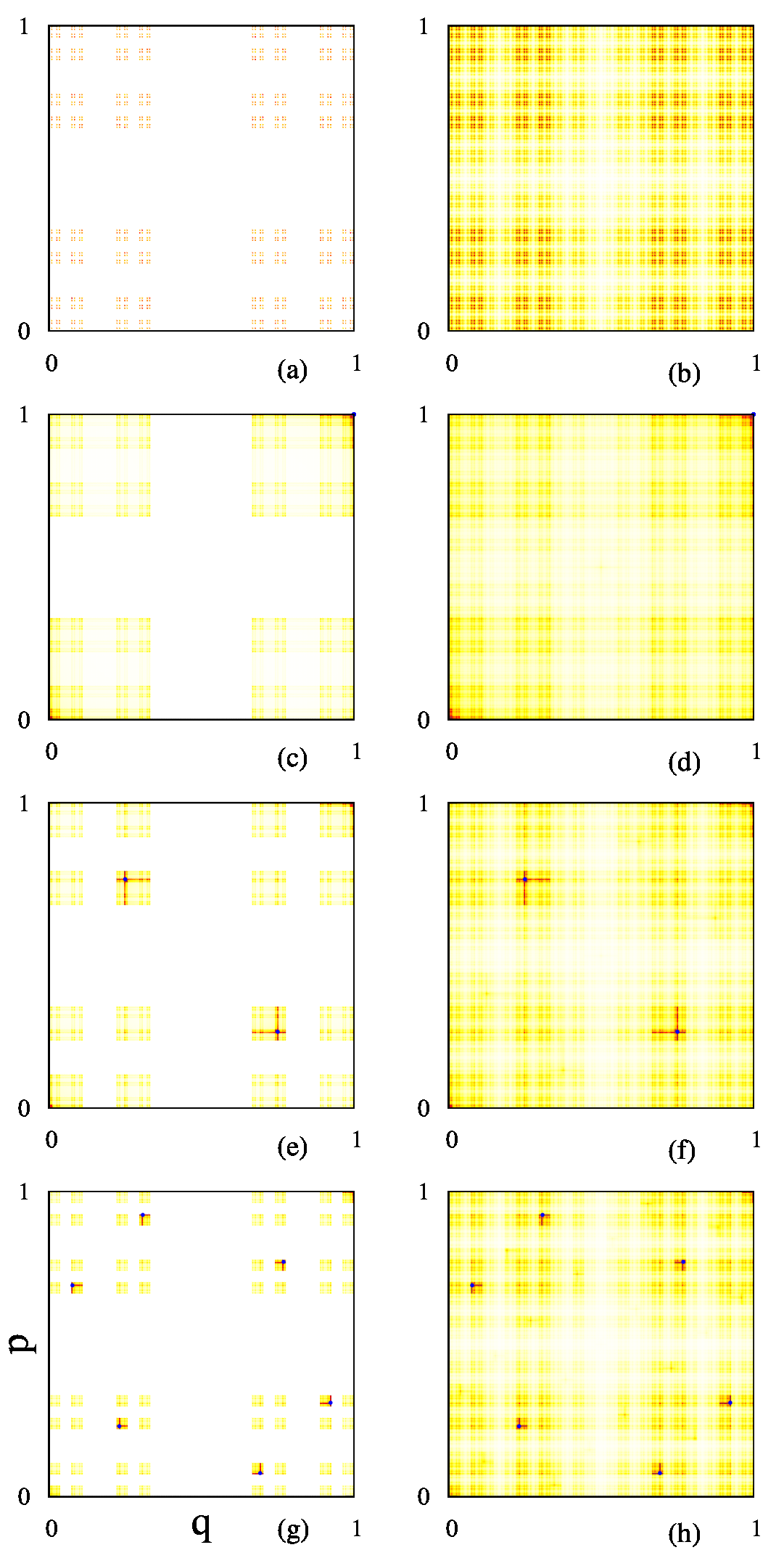}
\caption{(color online) Finite time repeller measure $\mu_{t,i}$ for the discontinuous 
reflectivity open tribaker map on the 2-torus at $t=10$ with $R=0$ (a) and $R=0.5$ (b). 
Corresponding values of the LDOs for the first (c)-(d), second (e)-(f), and third power 
(g)-(h) of the same map for $T=15$. The POs are marked by (blue) dark gray circles.
}
\label{fig1}
\end{figure}
\section{Results}
 \label{sec:results}

How does the description of the repeller by means of the measure $\mu$ compare with the LDOs? 
To answer this question we first examine the results of Fig.~\ref{fig1}.
In it, we show in the top row the finite time repeller measure $\mu_{t,i}$ 
at $t=10$ obtained for the constant reflectivity function, 
while in the second to fourth rows the corresponding LDOs for the first, second and third 
powers of the map with $T=15$ are displayed;
panels in the left and right columns correspond to $R=0$ and $R=0.5$, respectively.
Several comments are in order. First, it is clearly observed that the LDOs are peaked at 
the only surviving POs of period 1, 2, and 3 in the repeller, marked with (blue) dark gray circles 
in the Figure. Second, a substantial enhancement of the distribution 
around these POs is observed, while we do not find a significant difference with respect 
to this enhancement for the two values of the reflectivity, despite the fact that $R=0.5$ is 
only halfway to the closed map. Finally, what is left from the stable and unstable manifolds 
is clearly shown by the LDOs, whose values go to a smaller scale with growing period. 
What is more important, this suggests the way in which trajectories escape through the map opening.

%
\begin{figure}
\includegraphics[width=8.5cm]{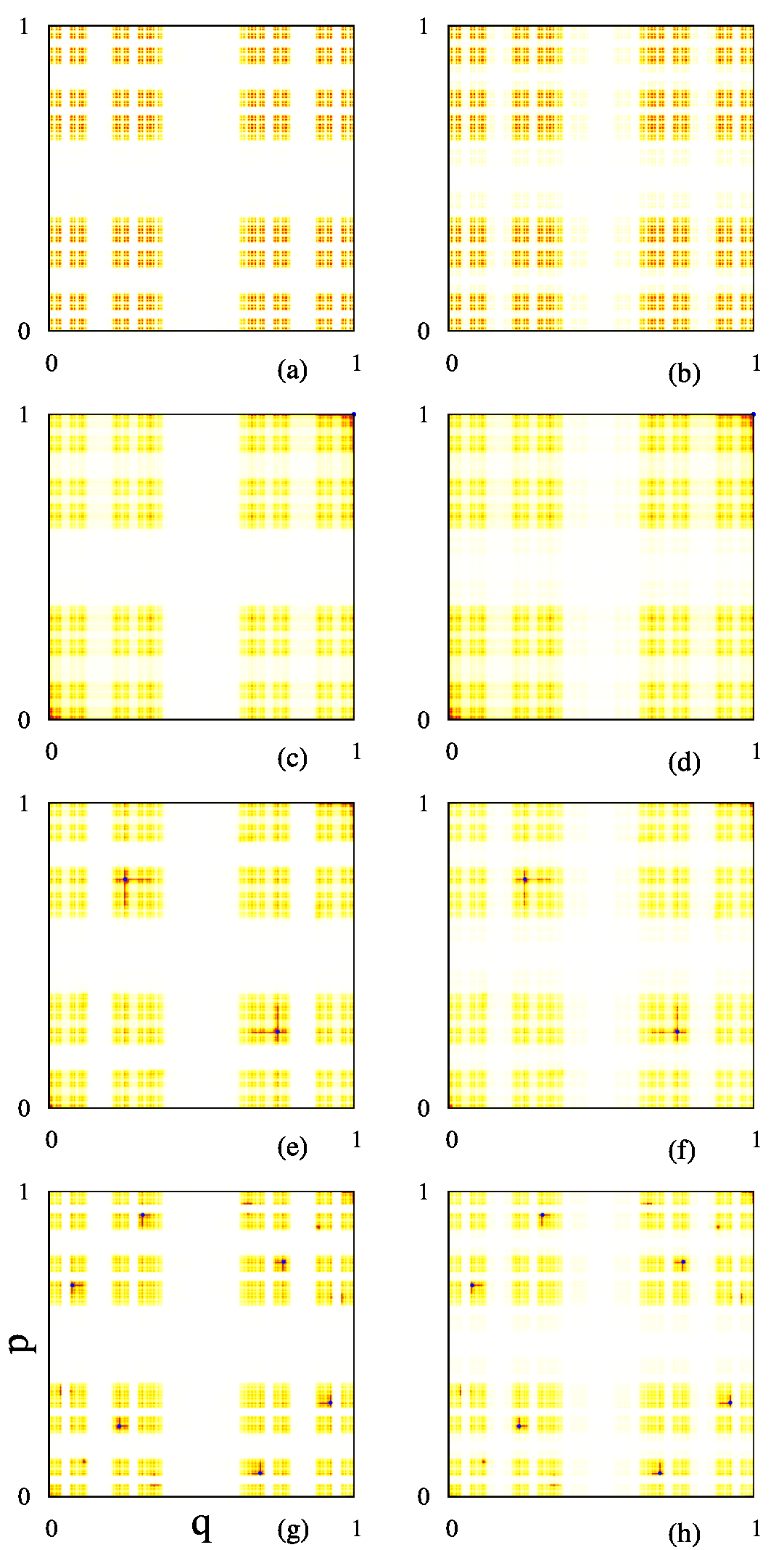}
\caption{(color online) Finite time repeller measure $\mu_{t,i}$ for the Fermi-Dirac type 
reflectivity open tribaker map on the 2-torus at $t=10$ with $R=0.01$ (a) and $R=0.1$ (b). 
Corresponding values of the LDOs for the first (c)-(d), second (e)-(f), and third power 
(g)-(h) of the same map for $T=15$. The POs are marked by (blue) dark gray circles.
}
\label{fig2}
\end{figure} 
In the case of the Fermi-Dirac type reflectivity
(see results in Fig.~\ref{fig2}), we obtain 
the same kind of behavior, namely the LDOs peak around the POs of each map, 
and there is essentially the same manifold structure for both $R=0.01$ and $R=0.1$. 
For this function, we have chosen two low reflectivities 
to compare how different the morphology of LDOs is in this regime. 
It is evident that though there is an order of magnitude difference 
between both $R$ values the map is essentially open with no significant discrepancies.

\begin{figure}
\includegraphics[width=9.5cm]{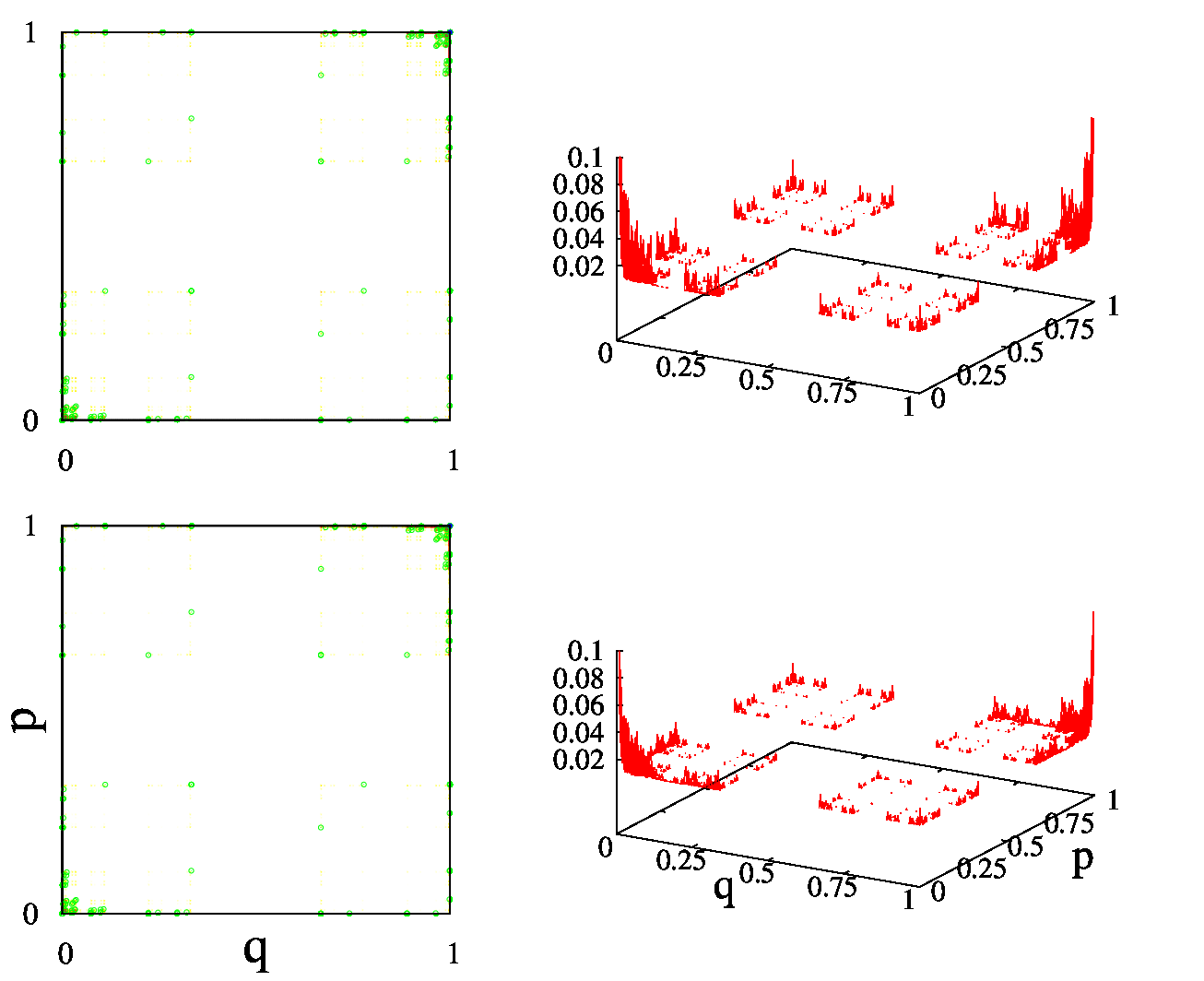}
\caption{(color online) LDO for the first power of the discontinuous opening tribaker 
map for $R=0$ (top), and $R=0.5$ bottom. 
On the left panels we show density plots together with the short times 
homoclinic tangle approximation with empty (green) light gray circles. 
The POs are marked by (blue) dark gray circles. 
On the right the corresponding three dimensional views are shown. 
In all cases data below 0.01 have been discarded.}
\label{fig3}
\end{figure}

\begin{figure}
\includegraphics[width=9.5cm]{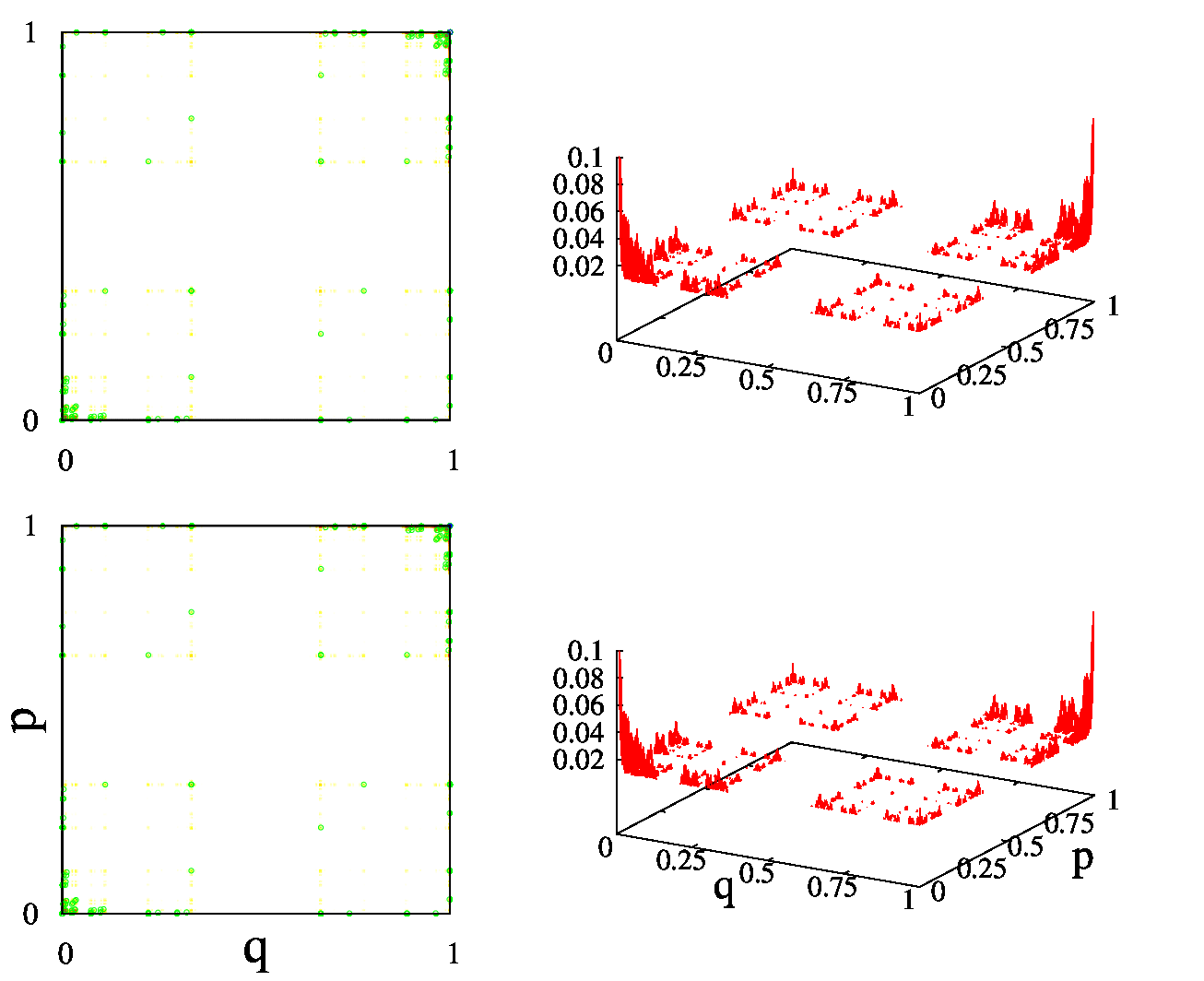}
\caption{(color online) LDO for the first power of the Fermi-Dirac type opening tribaker 
map for $R=0.01$ (top), and $R=0.1$ bottom. 
On the left panels we show density plots together with the short times 
homoclinic tangle approximation with empty (green) light gray circles. 
The POs are marked by (blue) dark gray circles. 
On the right the corresponding three dimensional views are shown. 
In all cases data below 0.01 have been discarded.
}
\label{fig4}
\end{figure}

We next analyze and compare the effect of both openings 
for different powers of the map. 
Looking at Figs.~\ref{fig3} and \ref{fig4} it is clear 
that the peaks are localized at the origin (or alternatively 
at the opposite corner identified with it in the torus), 
but also on a region around it. 
On the left columns, we can see the density plots, where too small 
values have been discarded for clarity. 
We plot superimposed the PO at $\{1,1\}$ ((blue) dark gray circles) and 
the short time approximation to the homoclinic tangle 
(empty (green) light gray circles) obtained with $e=3$. 
It is clear that the enhanced region 
agrees very well with the location of this tangle. 
In fact, if we look at the corresponding three dimensional views 
on the right columns a clear contrast between the values on the homoclinic 
tangle and the rest of the torus can be appreciated. 
It is remarkable that there is no big difference among the 
distributions for the discontinuous opening for $R=0.5$ and the Fermi-Dirac 
cases, though the completely open case shows more contrast.
%
%

\begin{figure}
\includegraphics[width=9.5cm]{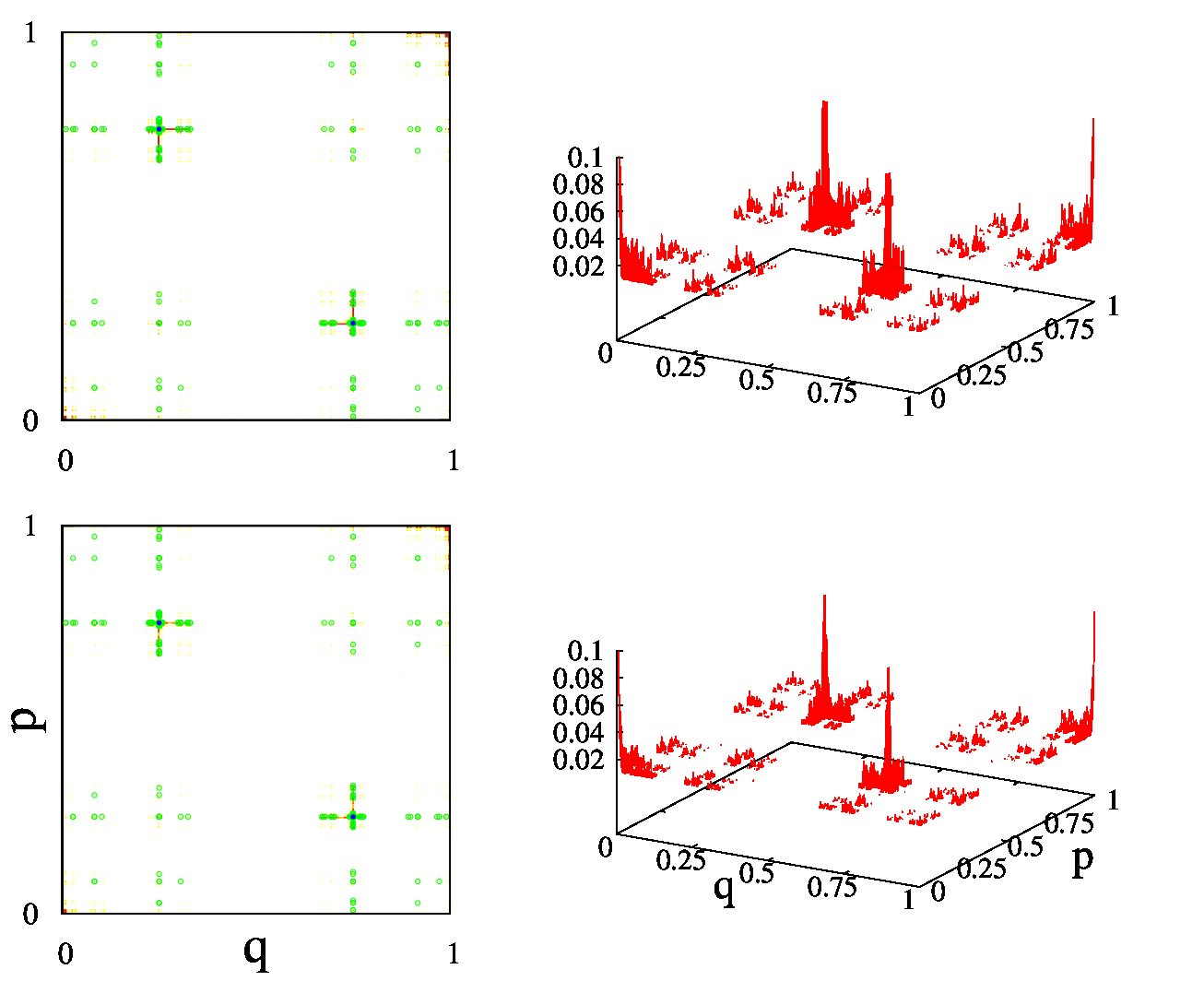}
\caption{(color online)  Same as Fig.~\ref{fig3} but for the second power 
of the corresponding map.
}
\label{fig5}
\end{figure}

\begin{figure}
\includegraphics[width=9.5cm]{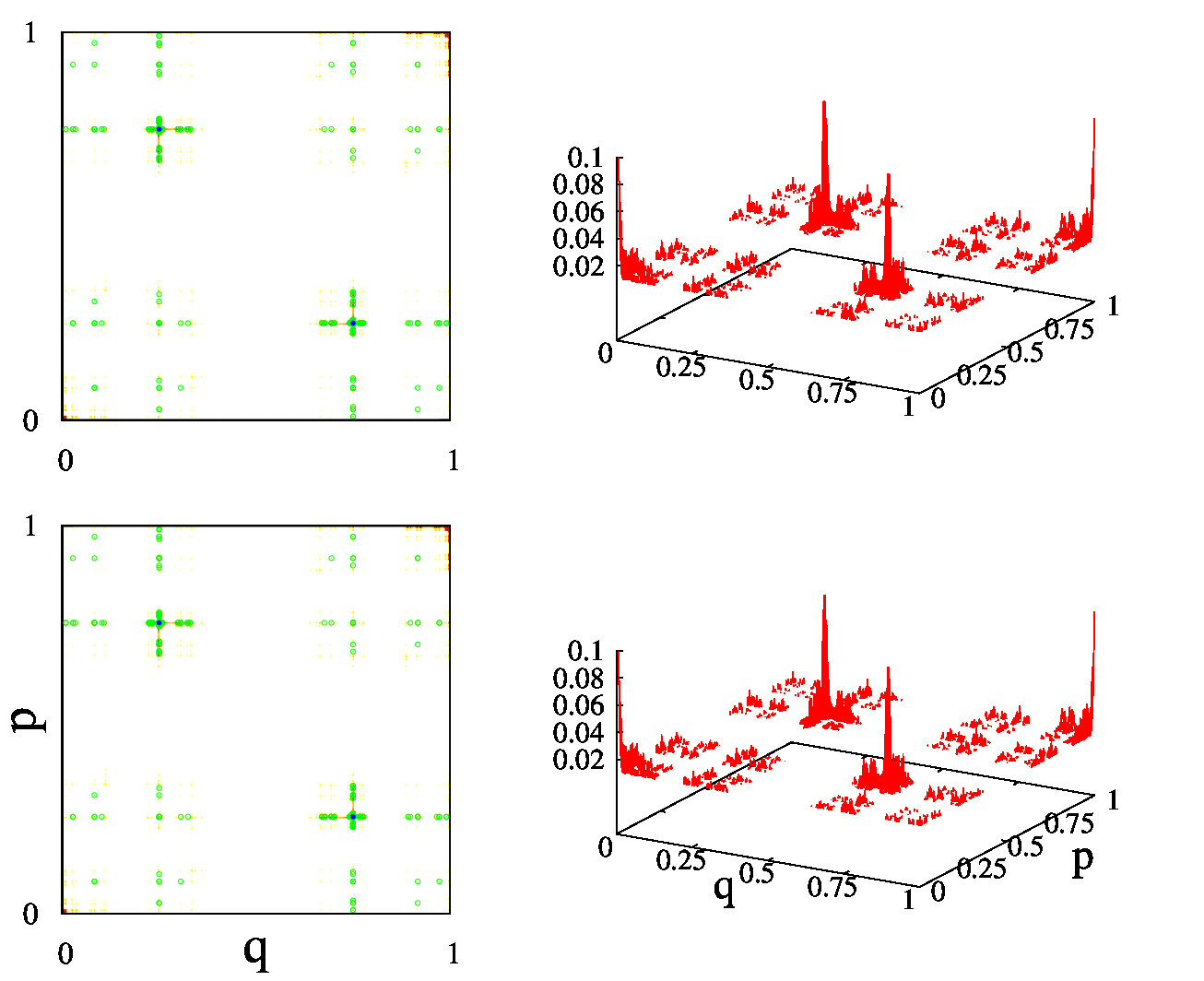}
\caption{(color online) Same as Fig.~\ref{fig4} but for the second power 
of the corresponding map.
}
\label{fig6}
\end{figure}

Results for the second power of the maps are shown in 
Figs.~\ref{fig5} and \ref{fig6}. 
The same behavior as in the case of the first power is found. 
Notice that the period 1 PO is a PO of this map so we find it again. 
Also, the only surviving period 2 orbit is clearly signaled by the LDOs, 
and moreover we can verify with the help of symbolic dynamics that its 
homoclinic tangle is very well described. 
Again, there is a higher contrast between the region associated with the 
short time approximation to the homoclinic tangle and the rest of the 
2-torus domain in the completely open scenario, i.e. the discontinuous opening 
case for $R=0$.
%
%

\begin{figure}
\includegraphics[width=9.5cm]{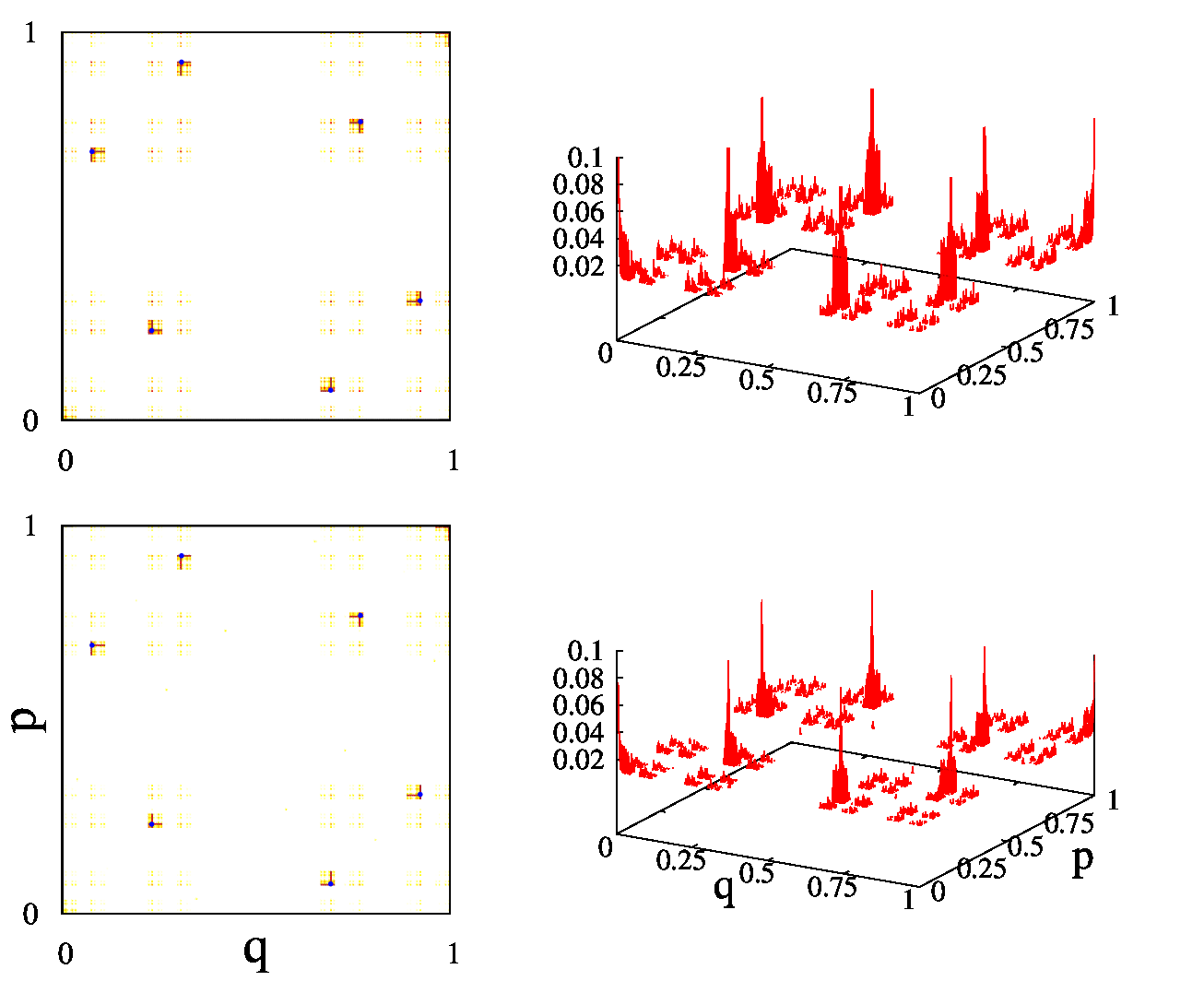}
\caption{(color online) Same as Fig.~\ref{fig3} but for the third power 
of the corresponding map (no homoclinic tangle approximation). 
}
\label{fig7}
\end{figure}

\begin{figure}
\includegraphics[width=9.5cm]{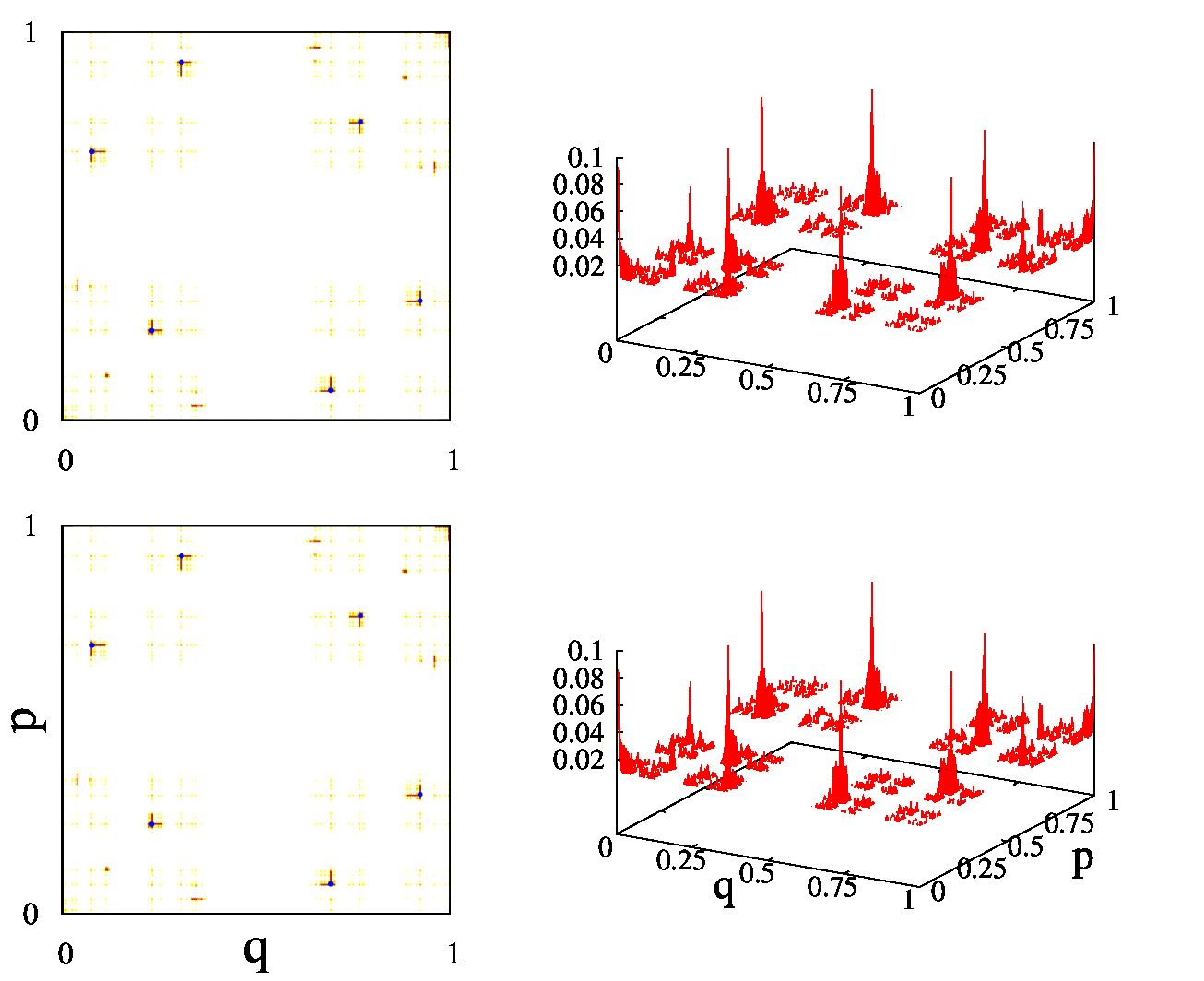}
\caption{(color online) Same as Fig.~\ref{fig4} but for the third power 
of the corresponding map (no homoclinic tangle approximation). 
}
\label{fig8}
\end{figure}

Finally, from Figs.~\ref{fig7} and \ref{fig8} it is clear that the third power 
of the maps shows the same behavior found in the two previous cases. 
Indeed, there is a clear enhancement of the LDOs around the surviving period 3 orbits. 
This time, we assume that it is the short time homoclinic tangle, 
but the calculations even with the help of symbolic dynamics become computationally 
difficult. However, this result shows the power of the LDOs in order to 
unveil the exact morphology of these sets in generic systems. 
We notice a small but interesting difference between the discontinuous case and 
the Fermi-Dirac one. 
Some peaks on POs inside the opening can be found and this could be ascribed 
to their location near the border of this region, and the smoothing of this 
boundary performed by the reflectivity function. 
We point out that this could give important information 
on the role played by the POs outside the repeller in the escape mechanism, 
with seemingly non trivial semiclassical consequences \cite{future}.
%
%

\begin{figure}
\includegraphics[width=8.5cm]{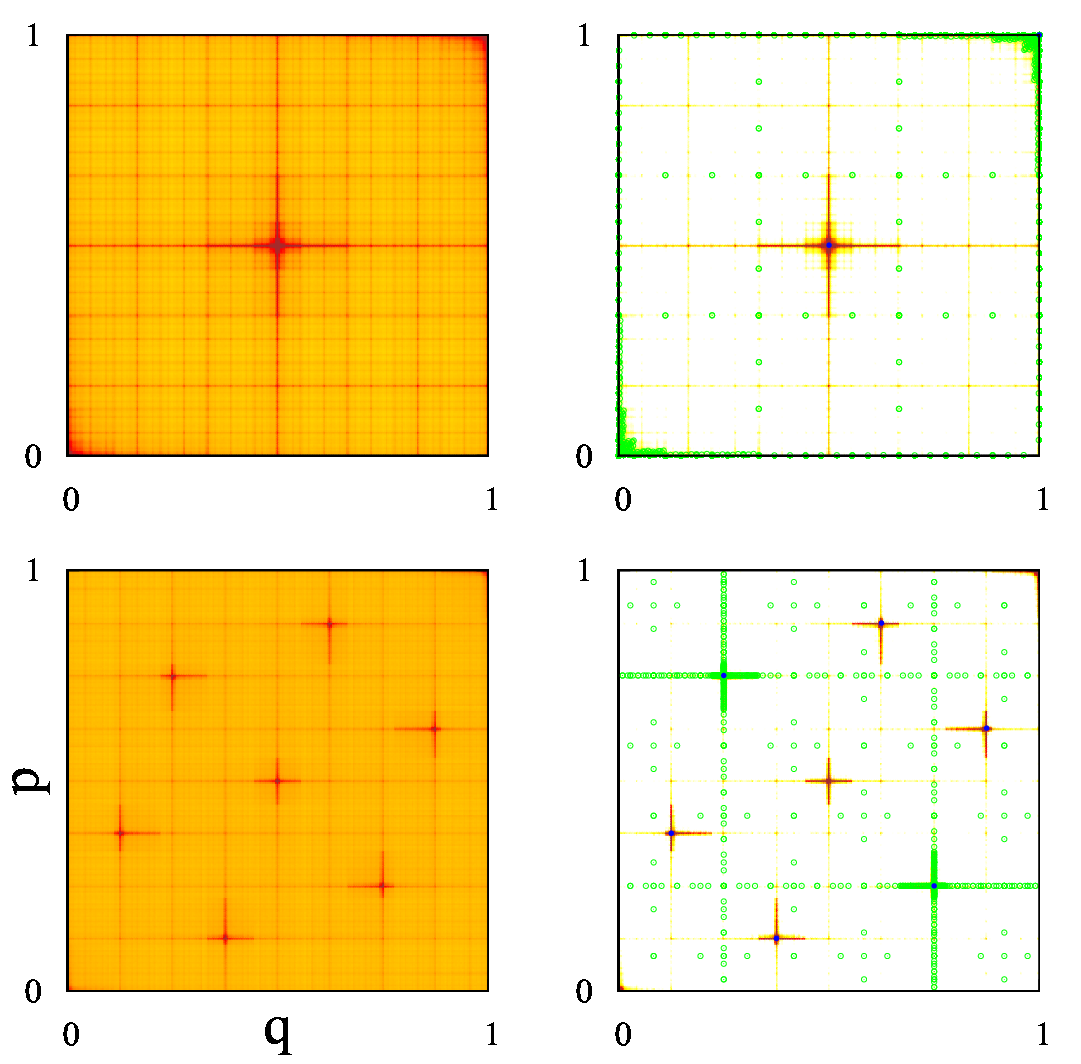}
\caption{(color online) LDOs for the first 
(top row) and second (bottom row) powers of 
the closed tribaker map. 
On the left we show the corresponding density plots considering 
all data values while, on the right we have discarded those below 0.0045. 
Also on the right we show the closed short times homoclinic tangle approximations of 
the same period 1 and 2 POs considered in the previous 
Figures with empty (green) light gray circles. 
The POs are marked by (blue) dark gray circles.}
\label{fig9}
\end{figure}

\begin{figure}
\includegraphics[width=9.5cm]{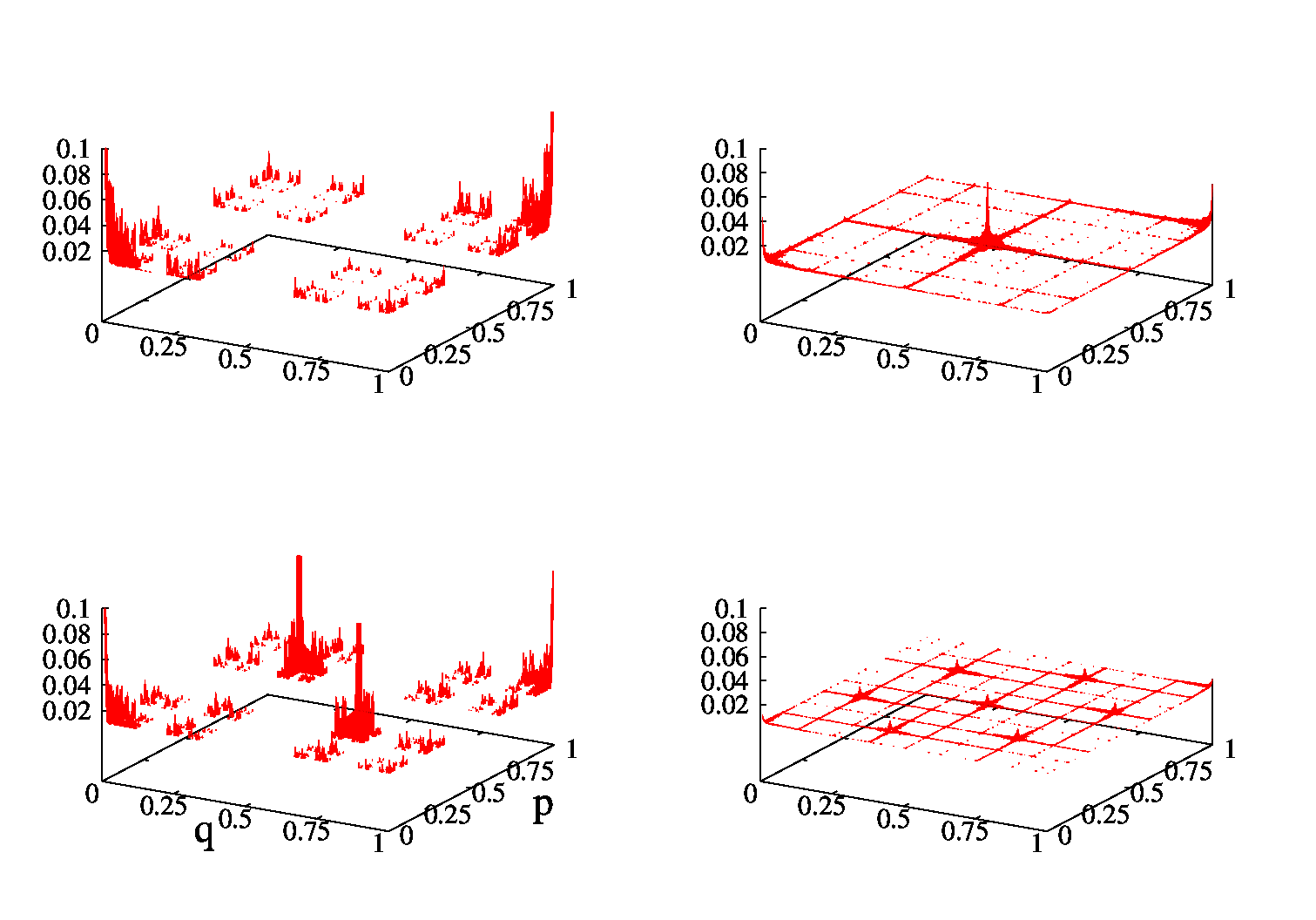}
\caption{(color online)  Three dimensional views of the LDOs for the first 
(top row) and second (bottom row) powers of the tribaker map. 
On the left column we show the results for the discontinuous opening case 
with $R=0$, and on the right one for the closed map ($R=1$).}
\label{fig10}
\end{figure}

What happens if we apply our definition of LDOs to the closed map? 
In principle, the motivation for adapting the original LDs was to 
unveil the inner structure of the repellers. 
But the LDOs definition when $I_t=1$ for all times can also be useful 
to locate the homoclinic tangle of POs of closed systems. 
In Fig.~\ref{fig9} we show the density plots for the first (upper row) 
and the second power of the closed map (lower row). 
On the left column all data values were taken into account and a clear 
enhancement around the POs can be appreciated. 
On the right column we discard the lowest values of LDOs and superimpose the 
short times approximation to the closed homoclinic tangle for the same 
two orbits of period 1 and 2 that survive in the repeller. 
Of course, now there are more orbits, but the description of the 
homoclinic sets is still valid.
%
However, if we look at Fig.~\ref{fig10} it becomes clear that the high contrast 
obtained for the open case is completely lost in the closed map. 
This result is due to the fact that the ratio of 
open to closed homoclinic orbits is given by $(2/3)^e$, 
which vanishes for $e \to \infty$. 
This makes LDOs to strongly peak on them for the open case, 
while in the closed scenario we have a more distributed situation. 
More importantly, this underlines in a very bold way the special suitability 
of the LDOs concept to uncover the details of repellers, including the 
overwhelming effect that any kind of opening has on its inner structure, 
specially on the remains of the homoclinic tangles.
%

\section{Conclusions}
 \label{sec:conclusions}
 
The recently introduced concept of LDs has proven to be very useful for unveiling 
the dynamical structures embedded in chaotic phase space. 
It is very amenable to numerical implementations and remarkably simpler than other 
methods for computational analysis of complex systems. 
Its definition in the discrete dynamical case has allowed to identify the stable 
and unstable manifolds associated to the POs of closed maps (autonomous and not), 
including the paradigmatic case of the Arnold's cat. 

We have successfully adapted this measure for open maps on the torus, defining the LDOs. 
With them, we have been able to identify the POs and the remains 
of the stable and unstable manifolds surviving in the repeller of the open
tribaker map in two reflectivity function cases, i.e.~the discontinuous and 
the Fermi-Dirac types of openings. 
This gives hints on possible applications for the study of trajectory escape 
mechanisms in more complicated systems. 
Moreover, with the aid of the very simple symbolic dynamics available for the 
tribaker map we could verify that our measure strongly peaks on the homoclinic 
tangles belonging to the POs. 
The high contrast between the values of the LDOs on these sets and the rest of 
the 2-torus not only allows for a very precise characterization of them, 
but it is indicative of the way in which the homoclinic circuits are truncated 
by the opening. 
It is important to notice that in both opening scenarios the structure of 
the repeller is essentially the same. 
This is an important result in that it provides a qualitative but unambiguous 
indicator of openness for a given map. 
As a matter of fact, this high contrast is almost completely lost in the closed case, 
where homoclinic orbits exponentially outnumber those belonging to the repeller.
We point out that these sets are not easy to identify in generic cases. 
Minor details associated to the kind of opening were also visualized, 
like for example the relevant role played in the escape from the repeller by
POs that do not belong to it when using the Fermi-Dirac type of reflectivity. 

All these results lead us to conclude that this work opens numerous possibilities 
for future research. On the one hand, chaotic scattering theory, where the 
homoclinic tangles play an essential role could receive new insights from LDOs 
characterization of generic complex systems. 
On the other hand, the semiclassical theory of short POs for open systems, 
that has raised the question of the role played by the POs outside of the repeller 
in the eigenfunctions description, could also greatly benefit from the use of LDOs 
to identify regions of higher relevance.
In fact, the search for scar functions contributions from these phase space 
regions could simplify the current POs selection criteria. 
 
\section{Acknowledgments}
Support from CONICET is gratefully acknowledged.
This research has also been partially funded by the Spanish Ministry of Science, 
Innovation and Universities, Gobierno de Espa\~na under Contract No.~PGC2018-093854-B-I00,
and by ICMAT Severo Ochoa under Contract SEV-2015-0554.

%
\end{document}